# REALIZAÇÃO DO MAPA LOGÍSTICO EM FPGA USANDO PADRÃO PONTO FIXO DE 32 BITS

Diego A. Silva[1], Eduardo B. Pereira[1], Erivelton G. Nepomuceno[1].

1. *Laboratório de Estudos em Sistemas Autônomos e Inteligentes, Departamento de Engenharia Elétrica,*
*Universidade Federal de São João del-Rei*
*Praça Frei Orlando, 170 – Centro – 36307-352 – São João del-Rei, MG, Brasil*
E-mails: `diego.augusto.silva95@gmail.com, ebento@ufsj.edu.br, nepomuceno@ufsj.edu.br`

***Abstract–*** *This article presents a design of the logistic map by means of FPGA (Field Programmable Gate Array) under fixed-point standard and 32-bits of precision. The design was carried out with Altera Quartus platform. The hardware description language VHDL-93 has been adopted and the results were simulated by means of Altera ModelSim package. The main of the project was to produce a chaotic system with a low energy and time cost. Using the VHDL, it was possible to use only 1439 logical gates from 114480 available. The Lyapunov exponent has been calculated with good agreement with literature reference, which shows the effectiveness the proposed method.*

***Resumo–*** Este artigo apresenta a implementação do mapa logístico usando FPGA (Field Programmable Gate Array) no padrão de 32 bits com ponto fixo. O projeto foi desenvolvido utilizando o software Quartus da Altera. Utilizou-se linguagem de descrição de hardware VHDL-93 e a simulação do resultado final foi feita no software ModelSim, também da Altera. O projeto teve como elemento norteador a proposição de um sistema caótico com baixo consumo de energia e tempo. Usando a linguagem VHDL, foi possível utilizar apenas 1439 portas lógicas de um total de 114480. O expoente de Lyapunov calculado e apresentou boa concordância com os valores presentes na literatura, confirmando o êxito do projeto.

**Palavras-chave–** FPGA, Mapa Logístico, Padrão Ponto Fixo, Modos de Arredondamento, Teoria do Caos.

## 1 Introdução

Sistemas caóticos são sistemas não-lineares determinísticos cujo comportamento dinâmico aparenta evoluir no tempo de maneira aleatória. Essa dinâmica complexa é extremamente sensível à variação de suas condições iniciais e paramétricas. Diversos sistemas de áreas do conhecimento distintas apresentam comportamento caótico, como por exemplo na Biologia, Ciências Sociais, Engenharia e Economia (May, 1976).

Na década de 1990, descobriu-se a possibilidade de sincronizar dois sistemas caóticos idênticos com condições iniciais diferentes, com aplicações em áreas como telecomunicações, simulações e criptografia (Pecora and Carroll, 1990). Desde então, uma parte da comunidade científica da área de sistemas caóticos dedica seus esforços no intuito de representar o comportamento caótico em sistemas eletrônicos.

As implementações de sistemas caóticos utilizando-se circuitos estão sujeitas a fatores como temperatura, umidade e envelhecimento de componentes que são capazes de gerar variações nos parâmetros destes sistemas. Para evitar esse tipo de problema, alguns pesquisadores passaram a utilizar implementações digitais dos mesmos, as quais são mais robustas com relação a variação de parâmetros. Dentre as implementações digitais, as implementações em *Field Programmable Gate Array* (FPGA) vêm surgindo como uma alternativa competitiva, uma vez que se tratam de dispositivos flexíveis e que consomem baixas quantidades de potência (Aseeri et al, 2002). Atratores de Chen (Sadoudi et al, 2009), Lü (Sadoudi et al, 2009) e Lorenz (De Micco and Larrondo, 2011) foram implementados por meio de um FPGA.

Estas implementações em meio digital vêm sendo estudadas em uma grande quantidade de trabalhos, tanto na construção de aplicações em hardware quanto em simulações computacionais. Porém, devido ao fato de esses sistemas utilizarem representações numéricas finitas as órbitas apresentadas são diferentes das obtidas via representação matemática infinita. Galias (2013) adverte sobre a necessidade de avaliar esses erros, e a propagação deles em funções recursivas pode ocasionar em uma grande perda de confiabilidade em resultados de simulações. Persohn & Povinelli (2016) alertaram sobre como desconsiderar as limitações provenientes de representações finitas em aplicações de criptografia pode levar à realização de sistemas inseguros. Alguns trabalhos, como Nepomuceno (2014) e Nepomuceno e Martins (2016), apresentam estratégias que podem ser utilizadas para avaliar o quanto uma simulação pode ser afetada por esses tipos de erros, criando métodos através dos quais pode-se rejeitar o resultado de uma simulação após certo ponto.

Uma das possíveis aplicações de sistemas caóticos está na geração de números aleatórios e criptografia (Dabal & Pelka, 2014), (Dabal & Pelka, 2011), (Pande & Zambreno, 2011). Com o crescimento do número de aplicações em sistemas embarcados, um aspecto que deseja-se investigar nesse trabalho é a obtenção de um sistema caótico realizado em FPGA que reduza consumo de energia e tempo realizado no processamento. Nesse sentido, pretende-se avaliar o comportamento do mapa logístico (May, 1976; Paiva et al, 2016), um dos mais conhecidos e investigados sistemas caóticos da literatura, utilizando um número menor de bits, 32, do que o usual 64 e arquitetura de padrão ponto fixo, ao invés de ponto flutuante. A representação em padrão ponto fixo em 32-bits apresenta menor utilização de área e baixo consumo de potência. Será analisada a resposta dessa arquitetura para condições conhecidas na literatura por apresentarem comportamentos caóticos utilizando dois diferentes modos de arredondamento e utilizando o expoente de *Lyapunov* como método para verificar se o sistema continua a apresentar caos. Todo o projeto foi desenvolvido utilizando o software *Quartus© II 17.0 Lite* da *Altera©*, sendo todo realizado utilizando linguagem de descrição de hardware VHDL-93. A arquitetura proposta será simulada através do software *ModelSim-Intel FPGA 10.5b*. O software *Matlab®* foi utilizado para a elaboração dos gráficos comparativos entre os diferentes resultados obtidos. A arquitetura desenvolvida possui como vantagens o fato de ser extremamente flexível, podendo facilmente ser modificada ou ter alguma melhoria adicionada, utilizar uma quantidade baixa de recursos de *hardware*, permitir um total controle sobre o fluxo de informação ao longo de todas as operações do sistema e representar o comportamento caótico do Mapa Logístico em uma representação de 32 bits em Padrão Ponto Fixo, ao invés de usar as representações em Ponto Flutuante de 64 bits utilizadas nas arquiteturas de processadores atuais e que consomem uma maior quantidade de área e potência.

## 2 Conceitos preliminares

### 2.1 Mapa Logístico

O mapa logístico é uma função recursiva que, para uma condição inicial $x_0$, fornece uma sequência numérica de saída de acordo com

$$x_{n+1} = rx_n(1 - x_n) \quad (1)$$

em que $n \in \aleph$ e $x_n$ e $r \in R$, de tal forma que $0 \leq x_n \leq 1$ e $0 < r \leq 4$ (May, 1976). À medida que $r$ se aproxima de 4 o comportamento do sistema vai se tornando caótico. Diferentes condicionamentos de (1), isto é, formulações matematicamente equivalentes, e formas diferentes de arredondamento podem levar a sequências numéricas diferentes para mesmos valores de $r$ e $x_0$ em representações finitas (Paiva et al, 2016). A equação (1) será implementada a partir de uma unidade que realiza o produto dos resultados das operações $(1 - x_n)$ e $(rx_n)$, realizadas em paralelo. O *hardware* possui configuração que suporta truncamento e arredondamento para o $+\infty$. Todas as operações serão realizadas em Padrão Ponto Fixo 32-bits.

### 2.2 Expoentes de Lyapunov

O expoente de *Lyapunov* é um número utilizado para analisar a sensibilidade de um sistema às suas condições iniciais. O cálculo do expoente de *Lyapunov* para uma função em tempo discreto é dado pela por

$$\lambda = \frac{1}{N}\sum_{i=0}^{N-1}|\ln f'(x_n)| \quad (2)$$

onde $N$ é a quantidade de elementos calculados para a série $f(x_n)$, e $f'(x_n)$ é a primeira derivada da função que representa a série. Para o Mapa Logístico, tem-se que $f(x_n)$ é dado pela equação (1) e $f'(x_n)$ é dado pela equação (3).

$$f'(x_n) = r(1 - 2x_n) \quad (3)$$

O valor de $r$ irá determinar o expoente de *Lyapunov* do Mapa Logístico. Para os casos em que $\lambda > 0$, os pontos da sequência se afastam exponencialmente da condição inicial, o que indica comportamento caótico (Monteiro, 2006).

### 2.3 Field Programmable Gate Arrays

Os *Field Programmable Gate Array* (FPGA) são dispositivos baseados em um arranjo contendo células lógicas e uma rede de interconexão que podem ser configurados para realizar funções específicas, cujas combinações podem sintetizar projetos com alto nível de complexidade. Essa configuração geralmente é feita através do uso de HDLs (*hardware description languages*). Muitos desses dispositivos possuem blocos de memória e circuitos multiplicadores embarcados. Nos últimos anos, os FPGAs aumentaram muito em capacidade, podendo conter bilhões de transistores. Devido à sua flexibilidade e a esse aumento de capacidade muitos projetistas vêm adotando os FPGAs em seus projetos, uma vez que os mesmos permitem a síntese de circuitos complexos a custos baixos e em curtos espaços de tempos, através do uso de softwares CAD oferecidos por fabricantes como a *Altera©* e a *Xilinx©* (Hamacher et al, 2012).

### 2.4 Ferramentas de programação e simulação voltadas para projetos em FPGA

Desenvolver projetos para FPGA é um processo que demanda transformações e algoritmos de otimização complexos. Com o intuito de automatizar alguns passos dessa tarefa, softwares com ferramentas específicas foram desenvolvidos pelos fabricantes de FPGA. O *Quartus II©* é um software desenvolvido pela *Altera©*, o qual contém uma coleção de ferramentas voltadas para projetos em FPGA. Dentre as funções que podem ser realizadas por meio dele pode-se destacar a construção de projetos e códigos com descrição em HDL, criação de modelos para simulação, análise de restrições temporais, compila-ção do projeto e configuração física do FPGA. O *ModelSim-Intel FPGA 10.5b* é um software desenvolvido pela *MentorGraphics©* cujo intuito é realizar simulações de descrições em HDL. Estas simulações permitem verificar se o hardware descrito funciona apropriadamente, atingindo os requerimentos especificados (Chu, 2011).

## 3 Metodologia

A seguir, a metodologia abordada no projeto é apresentada em quatro subseções, sendo que as três primeiras apresentam a implementação em FPGA e na última os métodos de avaliação dos resultados. O hardware desenvolvido foi dividido em duas unidades básicas: Uma unidade de controle, formada por um contador de iterações e uma máquina de estados finitos, e uma unidade operativa que implementa a equação (1). Todas as operações são realizadas em Padrão Ponto Fixo de 32 bits, onde são utilizados 16 bits para a parte inteira e 16 bits para a parte fracionária. O hardware possui configuração que suporta truncamento e arredondamento para o $+\infty$.

### 3.1 Unidade Operativa do Mapa Logístico

A parte principal da implementação em FPGA é mostrada na Figura 1. Nela temos uma unidade operacional principal denominada UOML (Unidade Operativa do Mapa Logístico). O principal objetivo dessa unidade é efetuar as operações em (1). Essa unidade contém sinais que indicam ocorrência de *overflow* e *underflow*, *o_over* e *o_under* respectivamente e o sinal *o_done*, que indica quando uma iteração completa de (1) é calculada. As entradas $i\_r$ e $i\_x_{n-1}$, recebem o coeficiente $r$ da equação (1) e o valor mais atualizado da sequência numérica, respectivamente. A entrada *i_round* é responsável por fixar o modo de arredondamento utilizado pelo sistema, sendo '1' para arredondamento para o $+\infty$ positivo e '0' para truncamento. O sinal denominado *i_ready* habilita o início da operação do bloco, sendo este controlado por uma máquina de estados finitos. Na Figura 1, também é mostrado o registrador $X_n$, que tem como função armazenar o valor atual de saída, $o\_x_n$ e realizar a realimentação para a próxima iteração da UOML. A implementação conta também com uma estrutura de roteamento por multiplexação, composta pelos elementos MUX1 e MUX2, os quais são controlados através dos estados da unidade de controle e pelo sinal *o_done*, sendo esta estrutura responsável por variar os valores na entrada do registrador $X_n$. Os sinais $i\_clk$ e $i\_rst$ representam o *clock* do sistema e uma entrada *reset* assíncrona, respectivamente. A cada novo cálculo realizado pela UOML a sua entrada $i\_x_{n-1}$ é atualizada com o valor calculado em sua iteração anterior, correspondente ao valor mais recente armazenado no registrador $X_n$.

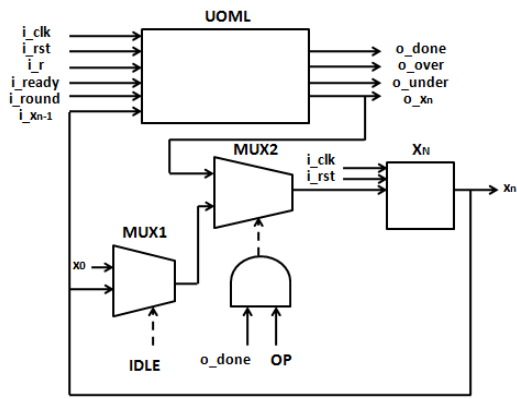

**Figura 1: Arquitetura da implementação em FPGA**

### 3.2 Unidade de Multiplicação e Conversão

A multiplicação aritmética entre dois números binários, com A e B bits, respectivamente, irá gerar um resultado com A + B bits (Hamacher et al, 2012). Para um sistema que realiza a multiplicação de dois números com 32 bits o resultado terá 64 bits. Isso não é desejável para um sistema que realiza operações em 32 bits e que está fazendo constante reuso de resultados de outras operações e iterações anteriores. Para evitar isso, foi adicionado uma unidade de conversão de 64 para 32 bits em série com cada operação de multiplicação. Essa unidade possui uma lógica combinacional cujo objetivo é gerar sinais que indiquem presença de *overflow* ou *underflow* durante a conversão. Caso haja *overflow* o número convertido será saturado e se houver *underflow* a conversão resultará em zero. Para melhor uso dos recursos da placa, que possui multiplicadores embarcados de 9-bits, as multiplicações foram realizadas através da decomposição dos números de 32 bits em partes iguais de 16 bits, cujos produtos foram computados em paralelo por *pipeline*. Essa configuração utiliza mais multiplicadores, porém melhora a performance do sistema com relação à frequência máxima de operação. Essas unidades multiplicadoras em série com os conversores são chamadas de Unidades de Multiplicação e Conversão. A UOML possui duas dessas unidades.

### 3.3 Unidade de Controle

Além da UOML, há também uma unidade de controle constituída por uma máquina de estados finitos, cujo fluxograma é apresentado na Figura 3, e um contador de 11 bits, cujo objetivo é determinar a quantidade de iterações do Mapa Logístico a serem realizadas, variando de 0 a 2047. A máquina possui quatro diferentes estados: *idle, op, done_it e done_all*. A Figura 2 mostra todos os estados, bem como todos os processos, decisões e transições relativos a cada um. Os retângulos escuros representam os estados, enquanto os mais claros representam processos e os losangos se relacionam às decisões. As transições são representadas pelas setas.

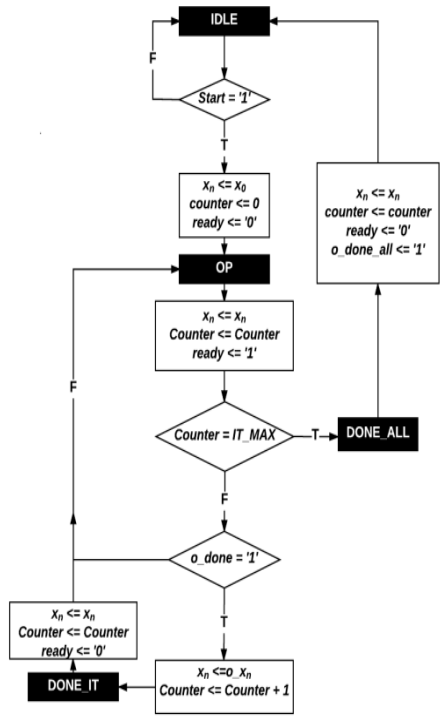

**Figura 2:** Fluxograma da máquina de estados finitos.

Durante o estado *idle*, a máquina se encontra inativa, esperando um sinal de entrada de *start* apresentar um nível lógico '1'. Nesse estado o registrador $X_n$ está conectado à condição inicial $x_0$, enquanto o contador *counter* e o sinal *ready* estão zerados. O sinal *ready* está ligado à entrada *i_ready* da UOML. Uma vez que o a condição *start* = '1' é satisfeita, há a transição para o estado *op*.

No estado *op* é realizada a operação da UOML. Caso o contador atinja seu valor máximo pré-estabelecido, *It_Max*, ocorre a transição para o estado *done_all*. Uma vez que o contador ainda não tenha alcançado seu valor máximo, toda vez que ocorrer um pulso do sinal *o_done*, que indica quando a UOML terminou sua operação, o registrador $X_n$ será conectado ao mais recente valor calculado pela UOML e a máquina irá para o estado *done_it*.

O estado *done_it* serve somente para realizar a transição entre duas iterações consecutivas. No estado *done_all* é gerado um pulso de meio ciclo de clock indicando que todas as iterações estão completas, seguido por uma transição ao estado *idle*.

### 3.4 Avaliação do hardware proposto

O *hardware* proposto foi avaliado em termos da utilização de recursos do FPGA, frequência máxima de operação e consumo de potência, os quais podem ser obtidos através do software *Quartus II©*, através de gráficos contendo as formas de onda dos sinais mais importantes do sistema, obtidos via simulação em VHDL-93 utilizando o software *ModelSim©*, e análise de um gráfico com a série temporal obtido via *Matlab®* utilizando dados exportados pela simulação. Serão calculados os expoentes de *Lyapunov* para os casos analisados, de acordo com a equação (2), e os mesmos serão comparados com valores já conhecidos da literatura. A frequência máxima de operação do *hardware* foi comparada com outras implementações presentes na literatura.

## 4 Resultados

Para realizar os testes com o hardware proposto foram adotados os valores $r = 4$ e $x_0 = 0{,}1$. O sistema foi configurado para realizar 150 iterações. Primeiramente, o projeto foi compilado utilizando o software *Quartus II©17.0 Lite* da *Altera©*. A Tabela 1 mostra a utilização de recursos físicos e a frequência de operação máxima do *hardware*, tomando como base o *FPGA Cyclone IV EP4CE115F29C7*, presente na placa *DE2-115*. Pode-se perceber que a quantidade de recursos utilizada é apenas uma pequena porção dos que estão disponíveis no FPGA. Com relação às restrições temporais o hardware proposto não apresentaria problemas ao ser implementado na placa, uma vez que a frequência máxima de operação é de 166,83 MHz, maior que a de 50 MHz disponível na placa.

**Tabela 1 –** Informações sobre o consumo de recursos e frequência máxima de operação do hardware proposto.

|  | Consumo do *hardware* | Recursos da Placa |
|---|---|---|
| Elementos lógicos | 1471 | 114480 |
| Multiplicadores de 9-bits | 16 | 532 |
| Registradores | 1358 | 114480 |
| Frequência máxima de operação | 166,83 MHz | 50 MHz |
| Bits de memória | 320 | 3981312 |
| Consumo de potência | 171,47mW | |

A Tabela 2 mostra um comparativo entre as frequências máximas de operação obtidas em implementações presentes na literatura e a obtida neste trabalho, junto com o número de bits utilizados por cada uma em Padrão Ponto Fixo. O presente trabalho foi desenvolvido em uma placa da *Altera©*, enquanto os demais foram desenvolvidos em placas da *Xilinx©*. Uma vez que a organização interna dos FPGAs desses fabricantes são diferentes não foram feitas comparações com relação aos recursos utilizados pelas implementações. Através da Tabela 2, pode-se perceber que a implementação do trabalho atual é competitiva com relação às demais, porém ainda pode ser melhorada com relação à precisão, uma vez que (Dabal & Pelka, 2014) e (Pande & Zambreno, 2011) utilizam 64 bits, e com relação à frequência máxima de operação, uma vez que (Dabal & Pelka, 2014) conseguiram um valor maior que o apresentado no presente trabalho.

**Tabela 2 –** Comparação entre frequências máximas de operação de diferentes implementações presentes na literatura.

| Trabalho analisado | Frequência máxima de Operação | Número de Bits |
|---|---|---|
| Atual | 166,83 MHz | 32 |
| (Dabal & Pelka, 2011) | 76,1 MHz | 32 |
| (Dabal & Pelka, 2014) | 233 MHz | 64 |
| (Pande & Zambreno, 2011) | 93 MHz | 64 |

Após a compilação do projeto, foram realizadas duas simulações utilizando o software *ModelSim-Intel FPGA 10.5b*. A primeira simulação foi realizada utilizando modo de arredondamento para o $+\infty$, enquanto a segunda utilizou do truncamento. As simulações geraram saídas sob a forma de arquivos de texto contendo informações relativas às 150 iterações adotadas. Esses arquivos foram processados via *Matlab*®, fornecendo as saídas gráficas presentes nas Figuras 3 e 4, as quais comparam as sequências numéricas para ambos os casos. A Figura 3 avalia as primeiras 50 iterações, ficando óbvio que após 20 iterações as duas órbitas já começam a apresentar consideráveis diferenças, o que pode ser explicado pela grande sensibilidade desses sistemas a uma pequena variação dos parâmetros causada por diferença nos modos de arredondamento. Na Figura 4, onde são mostradas 150 iterações, fica ainda mais nítida a influência do modo de arredondamento.

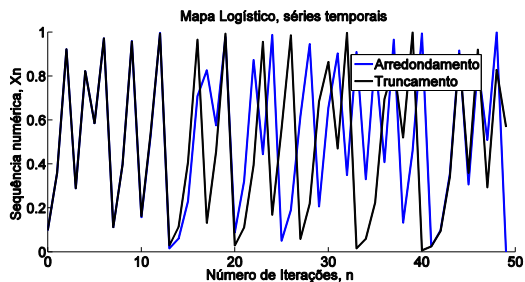

**Figura 3: Série temporal para as primeiras 50 iterações.**

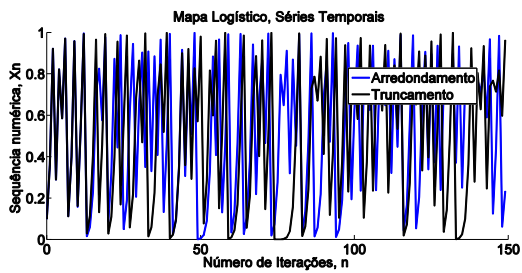

**Figura 4: Série temporal para as primeiras 150 iterações.**

A Tabela 3 mostra os expoentes de Lyapunov para os casos analisados. Para o caso em que $r = 4$, o expoente de *Lyapunov* tende a $\ln 2$, que é aproximadamente igual a 0,6931. Para o cálculo do expoente de Lyapunov foi utilizada a metodologia proposta em (Mendes e Nepomuceno, 2016) que permite o cálculo com um número menor de dados e é robusta quanto ao uso de números de bits inferior a 64. Os expoentes foram calculados para $N = 150$. Analisando-se a tabela pode-se perceber que os valores para ambas as séries se aproximam do valor esperado, logo o sistema reproduz o comportamento caótico do Mapa Logístico.

**Tabela 3 – Comparação entre o expoente de Lyapunov esperado pela literatura e os calculados através das séries obtidas.**

| Caso analisado | Expoente de *Lyapunov* |
|---|---|
| Literatura | 0,6931 |
| Série com arredondamento | 0,6979 |
| Série com truncamento | 0,7031 |

## 5 Conclusão

Neste artigo, é apresentada a implementação do Mapa Logístico em FPGA de modo a ilustrar a possibilidade de implementação de sistemas que apresentam dinâmica caótica em sistemas digitais. É apresentada a implementação da equação (1) para métodos de tratamento de casas decimais, arredondamento e truncamento. Os resultados mostram, por meio do expoente de *Lyapunov*, que a implementação em FPGA mostra a dinâmica caótica esperada pelo Mapa Logístico. Para cada método de representação das casas a série converge diferentemente ao longo do tempo, devido à sensibilidade dos sistemas caóticos às variações paramétricas. A arquitetura possui como vantagens o fato de consumir uma quantidade baixa dos recursos de *hardware*, consumindo pouco mais que 1% dos elementos lógicos do *FPGA* utilizado, ser extremamente flexível, podendo facilmente ser modificada ou melhorada, e permitir ao projetista um total controle sobre o fluxo de informações ao longo de todo o sistema. Para o futuro, espera-se expandir a realização do Mapa Logístico em FPGAs utilizando outras precisões numéricas e diferentes condicionamentos da equação, realizar a comunicação da arquitetura com outros dispositivos e testar aplicações práticas como a geração de números pseudoaleatórios.